\documentclass[pra,aps,showpacs,articles,nofootinbib,twocolumn]{revtex4-1}
\usepackage{epsf,epsfig}
\usepackage[psamsfonts]{amssymb}
\usepackage{amsmath}
\usepackage{color}

\def\be{\begin{equation}}
\def\ee{\end{equation}}
\def\ba{\begin{eqnarray}}
\def\ea{\end{eqnarray}}

\begin{document}

\title{Long-distance Entanglement generation by Local Rotational Protocols in spin chains}
\author{Morteza Rafiee}
\author{Hossein Mokhtari}
\affiliation{Department of Physics, Yazd University, Pajoohesh St,
Safaieh, 89195-741 Yazd, Iran}

\date{\today}

\begin{abstract}
We exploit the inherent entanglement of the ground state of a spin chain with dimerized $XX$ or $XXZ$ Hamiltonian to investigate the entanglement generation between the ends of the chain. We follow the strategy has been introduced in Ref. \cite{Abol-rotation} to encode the information in the entangled ground state of the system by local rotation. The amount of achieved entanglement in this scheme is higher than the attaching a pair of maximally entanglement scenarios. Also, our proposal can be implemented by using the optical lattices.
\end{abstract}

\pacs{03.65.Ud, 75.10.Pq, 03.67.Bg, 03.67.Hk}
\maketitle

\section{Introduction}
Entanglement lies at the heart of quantum mechanics and represents the most characteristic of it \cite{Schroedinger}. It is known to be key resource of quantum communication and computation \cite{QCC} and has been verified in such protocols as cryptography \cite{crypt} and teleportation \cite{telep}. Create a large amount of entanglement between distance subsystems is a much desire goal in quantum information tasks. One way to mediate interaction between distant qubits is to use an additional setup, called quantum bus. Spin chains are the most common buses where their tunable interaction has motivated researchers to use this permanently potential in the information processes \cite{bose,giovannetti,osborne,bayat3,bayat2}. Due to entanglement fragility under distance in most systems with short range interaction such as spin chains, one has to either delicately engineer the couplings \cite{Yung-bose,bayat-gate} or switch to super slow perturbative regimes \cite{Lukin-gate,WojcikLKGGB}. In all above studies symmetries of state and Hamiltonian seem play the important role versus inherent entanglement for propagating information \cite{Abol-rotation}. The entanglement inherent in many-body systems has been investigated \cite{Amico} and using it for a "known" state transferring has been recently proposed in Ref. \cite{Abol-rotation}. Sending a "known" state makes quantum communication simpler for many communication features such as key distribution \cite{crypt}. Furthermore, they have shown their proposal is more efficient to state transfer versus the previous scheme which attach a qubit encoding an "unknown" quantum state to the system \cite{bose2, bayat4}.

 The experimental realization of the Mott insulator phase for both bosons \cite{boson} and fermions \cite{fermion}, with exactly one atom, in optical lattices enables for realizing effective spin Hamiltonians \cite{Lukin} by properly controlling the intensity of laser beam . Moreover, Single qubit operations and measurements \cite{Meschede,Weitenberg,Gibbons-Nondestructive} are available by single site resolution in current experiments \cite{single-site}. Furthermore, singlet-triplet measurement of simulated spin have been done by using supperlattice\cite{supperlattice1,Trotzky1,Trotzky2,supperlattice2}.

In this letter, we put forward the approach in Ref.\cite{Abol-rotation} to investigate the amount of entanglement between the ends of a spin chain govern by $XX$ and $XXZ$ Hamiltonian. A setup consist of cold atoms trapped in a supperlattice has been introduced as a realization of our model.

The structure of this paper is as follows: in section (\ref{sec2}) we introduce our setup. In section (\ref{sec3}) entanglement generation is investigated with the chain where its dynamics govern by $XX$ Hamiltonian and in section (\ref{sec4}) the $XXZ$ Hamiltonian is considered and variation of obtainable entanglement in all the phase space is discussed. We finally summarize our results in section (\ref{sec5}).

\section{set-up}\label{sec2}

\begin{figure} \centering
    \includegraphics[width=8cm,height=5.5cm,angle=0]{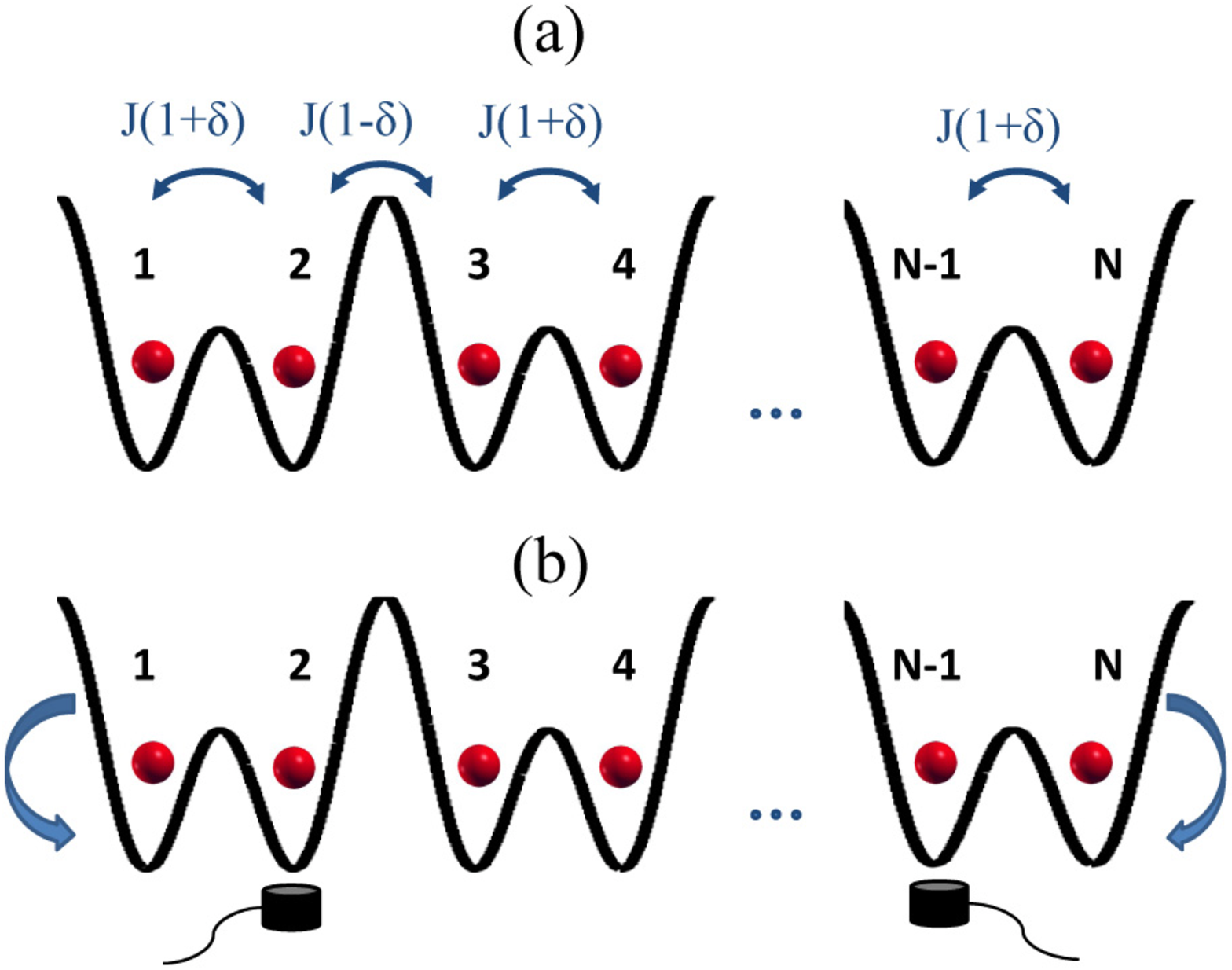}
    \caption{(Color online) (a) A superlattice containing $N$ atom and the tunnelings $J(1+\delta)$ and $J(1-\delta)$.
    (b) Encoding is done through local rotation on qubits $1$ and $N$, while decoding is done through measurement on qubits $2$ and $N-1$ at time $t^*$.}\label{fig1}
\end{figure}

We consider a chain of $N$ spin $1/2$ particles, where $N$ is even, interacting through a dimerized Hamiltonian
\begin{equation}\label{hamiltonian}
H=\sum_{j=1}^{N-1}(1+(-1)^{j+1} \delta)[J_x \sigma_j^x \sigma_{j+1}^x + J_y \sigma_j^y \sigma_{j+1}^y + J_z \sigma_j^z \sigma_{j+1}^z]
\end{equation}
where, $J_i>0$$(i=x,y,z)$ is the strength of coupling in the $i$ direction, $\sigma_j^i$$(i=x,y,z)$ denotes the Pauli operators at site $j$ and $0<\delta<1$ determines the dimerization of the chain. This interaction is called $XXZ$ and for $J_z=0$ is called $XY$ Hamiltonian and reduce to  Heisenberg Hamiltonian with $J_x=J_y=J_z=J$.
% As for the initial state of the chain ($\psi'(0)$) we consider the ferromagnetic state, with all the spins parallel to each other($|000...00 \rangle$), the state built as a series of singlets($|\psi^-,\psi^-,...\psi^- \rangle$), where $\psi^-=1/\sqrt(2)(|01\rangle-|10\rangle)$, and the ground state.
 We assume that the chain is in initial state and Alice controls qubits $1$ and $2$ while Bob controls the qubits $N-1$ and $N$.
 To encode two qubits at the ends of the chain Alice and Bob apply
 \begin{equation}\label{rotation}
 R_k(\theta,\phi)=\left(\begin{array}{*{20}c}
   {{\cos(\frac{\theta}{2})}} & {{-e^{-i\phi}\sin(\frac{\theta}{2})}} \\
   {{e^{i\phi}\sin(\frac{\theta}{2})}} & {{\cos(\frac{\theta}{2})}} \\
\end{array}\right).
\end{equation}
on first and last qubits ($k=1,N$) of the chain. After the operation $R$ the state of the chain changes to $|\psi(0)\rangle = R_1 R_N |GS\rangle$ 
%After encoding, 
and so the system evolves as $|\psi(t)\rangle=e^{-iHt}|\psi(0)\rangle$. At time $t=t^*$ the encoding state at each ends of the chain have been swapped while they are entangled via quantum gate at two middle qubits ($\frac{N}{2} and \frac{N+1}{2}$)\cite{bayat-gate,Lukin-gate}. Now Alice and Bob can localize this information in their single qubits by performing a single-qubit measurement in the computational basis on sites $2$ and $N-1$\cite{Abol-rotation}. % The qubits $2$ and $N-1$ can be projected on $|00\rangle$, $|01\rangle$, $|10\rangle$ or $|11\rangle$.
 After performing the projection amount of entanglement between the ends of the chain can be obtained by calculating the concurrence\cite{wooters}
\begin{equation}
C(t)=max\{0,2\lambda_{max}(t)-\sum_{i=1}^4\lambda_i(t)\},
\end{equation}
where $\lambda_i$ are the eigenvalues of the matrix $\sqrt{\rho_{1N} \sigma^y \otimes \sigma^y \rho_{1N}^* \sigma^y \otimes \sigma^y}$ while $\rho_{1N}$ is the reduced density matrix of the qubits $1$ and $N$.
As a physical realization of above Hamiltonian we propose the setting of ultracold atoms trapped in an optical supperlattice. An optical lattice mades of an standing wave formed by two different set of laser beams. The resulting potential is
\begin{equation}\label{V_lattice}
    V(x)=V_l \cos^2(2\pi x/ \lambda_l)+V_s \cos^2(2\pi x/\lambda_s)
\end{equation}
where, $\lambda_l=2\lambda_s$ are the wave lengths, $V_l$ and $V_s$ are the amplitudes. The low energy Hamiltonian of atoms trapped by $V(x)$
is \cite{Lukin}
\begin{align}\label{Hubbard}
   H=&-\sum_{<i,j>,\sigma} (J_{i\sigma} a^\dagger_{i,\sigma} a_{j,\sigma}+H.C.) + U_{\uparrow \downarrow} \sum_i n_{i,\uparrow}n_{i,\downarrow} \nonumber \\
   &+ \frac{1}{2}\sum_{i,\sigma}U_{\sigma}n_{i,\sigma}(n_{i,\sigma}-1) ,
\end{align}
where, $<i,j>$ denotes the nearest neighbor sites, $a_{i,\sigma}$ annihilates one atom with spin $\sigma=\uparrow,\downarrow$ at
site $i$, and $n_{i,\sigma}=a^\dagger_{i,\sigma} a_{i,\sigma}$. We are interested in the regime where $J_i\ll U_{\sigma},U_{\uparrow
\downarrow}$. This choice of hopping terms energetically prohibit the multiple occupancy of any site which corresponds to an
insulating phase. The effective Hamiltonian is found to be\cite{Lukin, clark}
\begin{eqnarray}\label{XXZ Hamiltonian}
   H&=&\sum_{<i,j>} J_i^{z}\sigma_i^z\sigma_j^z - \sum_{<i,j>} J_i^{\perp}(\sigma_i^x\sigma_j^x+\sigma_i^y\sigma_j^y),
\end{eqnarray}
where$\sigma_i^x=a^\dagger_{i\uparrow}a_{i\downarrow}+a^\dagger_{i\downarrow}a_{i\uparrow}$
and
$\sigma_i^y=-i(a^\dagger_{i\uparrow}a_{i\downarrow}-a^\dagger_{i\downarrow}a_{i\uparrow})$
are the pauli's spin operators. The effective couplings $J_i^z$ and $J_i^{ex}$ are given by
\begin{equation}
J_i^{z}=\frac{J_{i\uparrow}^2+J_{i\downarrow}^2}{2U_{\uparrow \downarrow}}-\frac{J_{i\uparrow}^2}{U_{\uparrow}}-\frac{J_{i\downarrow}^2}{U_{\downarrow}}, \hspace{10mm} J_i^{\perp}=\frac{J_{i\uparrow}+J_{i\downarrow}}{U_{\uparrow \downarrow}},
\end{equation}
The optical lattice parameters could be engineered such that $U_{\uparrow}=U_{\downarrow}=2U_{\uparrow\downarrow}=U$ and $J_{i\uparrow}=J_{i\downarrow}=J_i$. So the effective Hamiltonian reduced to the XX spin Hamiltonian \cite{Lukin, clark}.
\begin{eqnarray}\label{XX Hamiltonian}
   H&=&-\sum_{<i,j>} J_i^{\perp}(\sigma_i^x\sigma_j^x+\sigma_i^y\sigma_j^y),
\end{eqnarray}
 The tunneling $J_i$'s are controlled by the amplitudes $V_l$ and $V_s$\cite{Lukin}. Tuning the intensity of low(high) frequency trapping laser beam can be controlled independently the even (odd) couplings in a superlattice\cite{Trotzky1,Trotzky2}. So, it is possible to freeze the dynamics(J=0) at time $t^*$ with raising the barrier quickly and do the measurement on qubits.
 A schematic picture of the system is depicted in Fig.~\ref{fig1}(a) and (b).

\section{XX-model}\label{sec3}
In this section we consider a dimerized $XX$ model defined by Eq.(~\ref{hamiltonian}) with
\begin{equation}
J_x=J_y=\frac{J}{2}, \hspace{10mm} J_z=0.
\end{equation}
So the Eq.(~\ref{hamiltonian}) reduced to
\begin{equation}\label{xx-hamiltonian}
H=\frac{J}{2}\sum_{j=1}^{N-1}(1+(-1)^{j+1} \delta)[\sigma_j^x \sigma_{j+1}^x + \sigma_j^y \sigma_{j+1}^y]
\end{equation}
 This Hamiltonian can then be diagonalized with following the procedure described in\cite{Leib} with $\delta=0$. The first step is to perform a Jordan-Wigner transformation\cite{jordan-wigner} $c_l=\prod_{j=0}^{l-1}(-\sigma_j^z)\sigma_l$ where $\sigma^{\pm}=(\sigma^x \pm \sigma^y)/2$, $\{c_l,c_{l'}\}=0$ and $\{ c_l, c_{l'}^\dagger \}=\delta_{l,l'}$. As a result, the Hamiltonian is mapped in the free fermion Hamiltonian
\begin{equation}\label{fermion-xy}
H=J\sum_{j=1}^{N-1} (c_j^\dagger c_{j+1}+c_{j+1}^\dagger c_j)={\bf c}^\dagger {\bf M} {\bf c},
\end{equation}
where ${\bf c}=(c_1,...,c_N)$(${\bf c}^\dagger$) is the vector of the $N$ creation(anihilation) operators, and ${\bf M}$ is the adjacency matrix. The new fermionic operators are defined as $c_j=\sum_k \xi_k^i c_k$ with the eigenvalues $\lambda_k$ and eigenvectors $\xi_k$ of the matrix ${\bf M}$. So the Hamiltonian (~\ref{fermion-xy}) takes the form\cite{venuti}
\begin{equation}
H=\sum_k \lambda_k c_k^\dagger c_k.
\end{equation}
Also, dimerized XX Hamiltonian ($\delta \neq 0$) can be diagonalize with the procedure described in \cite{dimer_xy} for the odd number of qubits and in \cite{even-xy} for the chain with even number of spins. The eigenvalues $\lambda_k$ have been introduced in appendix for the nonvanishing amount of $\delta$.
 With this dimerized Hamiltonian we have a Werner state $\rho_w=p|\psi^- \rangle \langle \psi^-|+(1-p)\hat{I}_4/4$ for reduced density matrix of the first or last two qubits\cite{Abol-rotation}, where $|\psi^- \rangle=(|01 \rangle -|10 \rangle)/\sqrt{2}$, $\hat{I}_n$ is an $n \times n$ identity matrix and $p \rightarrow 1$ if $\delta \rightarrow 1$.
 %%%%%%%%%%%%%%%%%%%%%%%%%%%%%%%%

\begin{figure} \centering
    \includegraphics[width=8cm,height=5.5cm,angle=0]{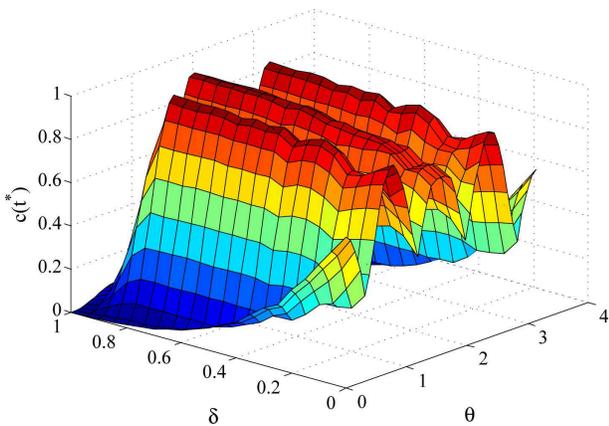}
    \caption{(Color online) Variation of the entanglement at $t^*$ versus $\theta$ and $\delta$ for a XX spin chain with N=8 and $\phi=0$.}\label{theta}
\end{figure}

\begin{figure} \centering
    \includegraphics[width=8cm,height=5.5cm,angle=0]{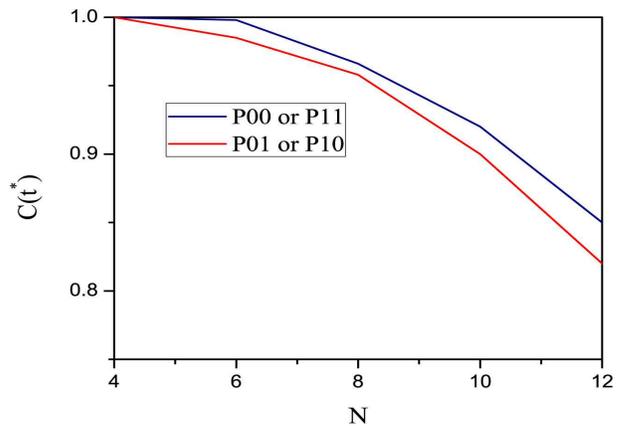}
    \caption{(Color online) The variation of entanglement vs. the number of qubits for a $XX$ spin chain with $\delta=0.8$ by applying different projection on qubits $2$ and $N-1$.}
     \label{proj}
\end{figure}

 The projection operators of two qubits have the forms $P00=|00\rangle \langle 00|$, $P01=|01\rangle \langle 01|$, $P10=|10\rangle \langle 10|$ and $P11=|11\rangle \langle 11|$, where $P00$ means both of two qubits are projected on $|0 \rangle$. Alice and Bob apply these projections on qubits 2 and $N-1$. At the first step, the best choice of $\theta$ and $\delta$ should be determined. So the variation of the entanglement at $t=t^*$ between the ends of the chain with $N=8$ and $\phi=0$ versus $\theta$ and $\delta$ has been calculated while the projection $P00$ or $P11$ has been applied. These results which have been plotted in Fig. (\ref{theta}) show the best amount of $\theta=\pi/2$ and $\delta=0.8$. The concurrence decrease for $\delta>0.8$ due to emergence of small couplings(i.e. $J(1-\delta)$). Entanglement at time $t^*$ between ends of the chain with $N=10$ and $\delta=0.8$ has been plotted in Fig.~\ref{proj} for different projection operators. So, performing $P00$($P11$) is the best choice and we use this projection in the following calculations.

\begin{figure} \centering
    \includegraphics[width=8cm,height=5.5cm,angle=0]{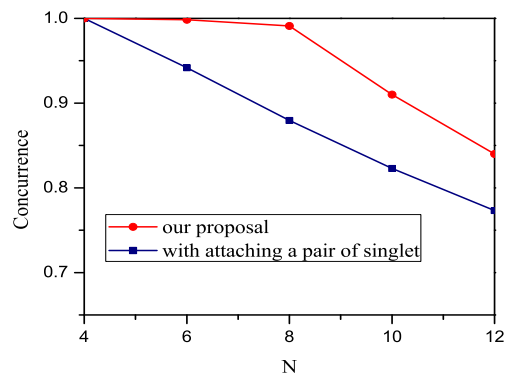}
    \caption{(Color online) Comparison between different strategies of entanglement generation between the ends of a $XX$ spin chain namely, entanglement achieved in our scheme(red line) and entanglement obtained with attaching an extra maximally entangled pair(blue line).}
     \label{xx-compare}
 \end{figure}

  Furthermore, we compare the amount of entanglement achieved in our proposal and anti-ferromagnetic chain with attaching a pair of maximally entanglement\cite{bayat4} in Fig~\ref{xx-compare}. This figure shows that the amount of entanglement in our proposal is higher than attaching scheme and it is in agreement with the compare of the amounts of average fidelity for state transfer in Ref \cite{Abol-rotation}.

\section{XXZ Hamiltonian}\label{sec4}
\begin{figure} \centering
    \includegraphics[width=8cm,height=5.5cm,angle=0]{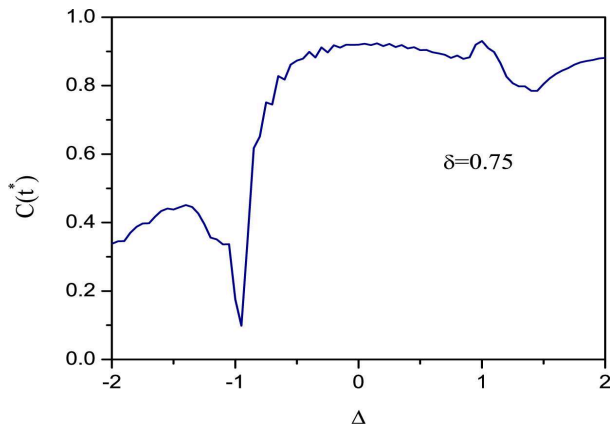}
    \caption{(Color online)Variation of entanglement between the ends of the "XXZ" chain of $N=10$ as a function of $\Delta$ with dimerized parameter $\delta=0.75$.}
     \label{XXZ-phase}
\end{figure}

\begin{figure} \centering
    \includegraphics[width=8cm,height=5.5cm,angle=0]{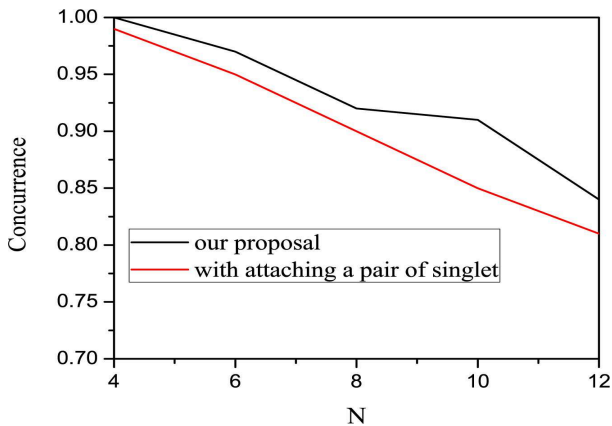}
    \caption{(Color online)Comparison between different strategies of entanglement generation between the ends of a $XXZ$ spin chain with $\Delta=0.5$ namely, entanglement achieved in our scheme(black line) and entanglement obtained with attaching an extra maximally entangled pair(red line).}
     \label{xxz-compare}
\end{figure}

In this section we consider a dimerized $XXZ$ model defined by Eq.(~\ref{hamiltonian}) with
\begin{equation}
J_x=J_y=\frac{J}{2}, \hspace{10mm} J_z=\frac{J \Delta}{2}.
\end{equation}
where $\Delta$ is the anisotropy coupling in the $z$ direction.
The above Hamiltonian is a dimerized $XXZ$ Hamiltonian. The usual $XXZ$ Hamiltonian has a very rich phase diagram which different phases depend on different range of $J$ and $\Delta$.
For $\Delta$ = 1 and $J < 0$, this interaction is the FM Heisenberg chain widely discussed in the context
of quantum communication \cite{bose,bose2,xxz-phase1}. More interesting regimes exist for $J > 0$ and different values of $\Delta$ \cite{xxz-phase2}. $\Delta < −1$ is
the FM phase with a simple separable biased ground state with all spins aligned to the same direction. $−1 < \Delta \leq 1$ is
called XY phase, which is a gapless phase and consists of two different legs, the FM half ($−1 < \Delta < 0$) and the AFM part
($0 \leq \Delta \leq 1$). $1 < \Delta $ is called N\'{e}el phase. In the Ising limit $\Delta \gg 1$ the ground state is the N\'{e}el state $(|010101 . . . 01 \rangle)$.
We use the same recipe for entanglement generation with this Hamiltonian.
The concurrence between qubits $1$ and $N$ for the chain with $N=10$ at time $t^*$ has been plotted in the domain of ($-2 \leq \Delta \leq 2)$
in Fig.~\ref{XXZ-phase} with $\delta=0.75$ as a best amount of dimerized parameter for this Hamiltonian. As we can see in this figure for our proposal works only for the domain $-1 < \Delta$. The reduced density matrix of first or last two qubits is the Werner state in this domain of phase space. On the other side, there is no local entanglement at ground state for the transition point $\Delta=-1$ and the domain $\Delta<-1$ so, our mechanism which exploit inherent entanglement between proximally spins in ground state doesn't work in this region. Also, obtainable entanglement between ends of a chain with different length $N$ is enhanced compared to entanglement achieved by attaching a pair of maximally entanglement to the system while $\Delta=0.5$ as has been shown in Fig.~\ref{xxz-compare}

\section{Conclusion}\label{sec5}
 In this paper we examined inherent entanglement in ground state of the spin chains for investigation of entanglement generation between the ends of chain. Dynamics of the chains govern by dimerized $XX$ and $XXZ$ Hamiltonian which can be realized by optical supperlattice. In this scheme local rotation on the ends of the chain encode information in the entangled ground state of the system. We also showed that the obtainable entanglement is higher than entanglement achieved by attaching a pair of maximally entanglement to the system. For $XXZ$ Hamiltonian this mechanism doesn't work for the domain with $\Delta\leq -1$. 
\section{Acknowledgments}
Authors thank A. Bayat for useful discussion and comments at
university of Ulm.

\appendix
\section{Eigenvalues of dimerized XX Hamiltonian}
Upon the introduction of a nonvanishing $\delta$ the eigenvalues of the adjacency matrix ${\bf M}$ for odd $N$ are given by \cite{dimer_xy, even-xy}
\begin{equation}
\lambda_k=\left \{
 \begin{array}{cc}
 J(1+\delta)\sqrt{\Delta_k} &\quad{\rm for}\quad k=1,2,...\frac{N-1}{2} \\
 0 &{\rm for}\quad \quad \quad k=\frac{N+1}{2}\\
 -J(1+\delta)\sqrt{\Delta_k} &\quad{\rm for}\quad k=\frac{N+3}{2},\frac{N+5}{2},...
  \end{array}
  \right.,
 \end{equation}
 where
 \begin{equation}
 \Delta_k=1+2\frac{1-\delta}{1+\delta}\cos(\frac{2\pi k}{N+1})+(\frac{1-\delta}{1+\delta})^2.
 \end{equation}
The eigenvalues of adjacency matrix ${\bf M}$ for even $N$ are given by \cite{even-xy}
\begin{equation}
\lambda_k=\left \{
 \begin{array}{cc}
 J(1+\delta)\sqrt{\Delta'_k} &\quad{\rm for}\quad k=1,2,...\frac{N-1}{2} \\
  -J(1+\delta)\sqrt{\Delta'_k} &\quad{\rm for}\quad k=\frac{N+3}{2},\frac{N+5}{2},...
  \end{array}
  \right.,
 \end{equation}
 where
 \begin{equation}
 \Delta'_k=1+\frac{1-\delta}{1+\delta}\cos(x_{\nu})+(\frac{1-\delta}{1+\delta})^2.
 \end{equation}
 and $x_{\nu}$ are the solutions of equation($\frac{1-\delta}{1+\delta}\sin(\frac{N}{2}x_{\nu})+\sin((\frac{N}{2}+1)x_{\nu})=0$) which has different solution for $\frac{1-\delta}{1+\delta}<(N+2)/N$ and $\frac{1-\delta}{1+\delta} \geq (N+2)/N$.

%\newpage

%\begin{figure} \centering
%    \includegraphics[width=8cm,height=5.5cm,angle=0]{xx-delta.eps}
%    \caption{(Color online) Entanglement at time $t^*$ as a function of dimerization parameter $\delta$ for a chain of $N=10$.}
%     \label{xx-delta}
%\end{figure}


\begin{thebibliography}{99}
\bibitem{Schroedinger}
E. Schr\"{o}dinger, Proc. Cambridge Phil. Soc., {\bf 31}, 555 (1935);
{\bf 32}, 446 (1936).

\bibitem{QCC}
M. Nielsen, I. Chuang, {\it Quantum Information and Quantum Computation} (Cambridge University Press, Cambridge, 2000).

\bibitem{crypt} N. Gisin, G. Ribordy, W. Tittel, and H. Zbinden, Rev. Mod.
Phys. {\bf 74}, 145 (2002).

\bibitem{telep} D. Bouwmeester, J.-W. Pan, K. Mattle, M. Eibl, H. Weinfurter,
and A. Zeilinger, Nature (London) {\bf 390}, 575 (1997); D. Boschi, S. Branca, F. De Martini, L. Hardy, and S. Popescu,
Phys. Rev. Lett. {\bf 80}, 1121 (1998).

\bibitem{bose}S. Bose, Phys. Rev. Lett. {\bf 91}, 207901 (2003); J. Eisert {\em et. al.}, Phys. Rev. Lett. {\bf 93}, 190402 (2004); M.~Christandl {\em et. al.},Phys. Rev. Lett. \textbf{92}, 187902 (2004).

\bibitem{giovannetti}V. Giovannetti and D. Burgarth, Phys. Rev. Lett. {\bf 96}, 030501
(2006); J. Fitzsimons and J. Twamley, Phys. Rev. Lett. {\bf 97},
090502 (2006); A. Kay, Phys. Rev. Lett. {\bf 98}, 010501 (2007).

\bibitem{osborne}T. J. Osborne and N.~Linden, Phys. Rev. A \textbf{69}, 052315
(2004);  A. Lyakhov and C. Bruder, Phys. Rev. B {\bf 74}, 235303
(2006).

\bibitem{bayat3}A. Bayat and V. Karimipour, Phys. Rev. A {\bf 71}, 042330
(2005). D. Burgarth, S. Bose, Phys. Rev. A 73, 062321 (2006); L. Zhou, J. Lu, T. Shi and C. P. Sun,
quant-ph/0608135. D. Burgarth and S. Bose, Phys. Rev. A {\bf 71}, 052315
(2005); M. Avellino, A. J. Fisher, S. Bose, Phys. Rev. A {\bf 74},
012321 (2006).

\bibitem{bayat2}A. Bayat and S. Bose, Advances in Mathematical Physics, \textbf{2010}, 127182
(2010); A. Bayat, D. Burgarth, S. Mancini and S. Bose, Phys. Rev.
A \textbf{77}, 050306(R) (2008).

\bibitem{Yung-bose} M.~.H.~Yung and S.~Bose, Phys. Rev. A {\bf 71}, 032310 (2005);
M.~.H~Yung, S.~C.~Benjamin and S.~Bose, Phys. Rev. Lett. {\bf 96},
220501 (2006).

\bibitem{bayat-gate} L.~Banchi, A.~Bayat, P.~Verrucchi, and S.~Bose, Rev. Lett. {\bf 106}, 140501 (2011).

\bibitem{Lukin-gate} N.~Y. Yao {\em et~al.}, Phys. Rev. Lett. {\bf 106}, 040505 (2011);
N.~Y. Yao {\em et~al.}, arXiv:1012.2864.

\bibitem{WojcikLKGGB} A.~{W{\'o}jcik} {\em et~al.}, Phys. Rev. A {\bf 72}, 034303 (2005).

\bibitem{Abol-rotation} S. Yang, A. Bayat and S. Bose, Phys. Rev. A {\bf 84}, 020302(R) (2011).

\bibitem{Amico} L. Amico, R. Fazio, A. Osterloh and V. Vedral, Rev. Mod. Phys. {\bf 80}, 517 (2008).
\bibitem{bose2} S. Bose, Contemporary Physics {\bf 48}, 13 (2007).
\bibitem{bayat4} A. Bayat and S. Bose, Phys. Rev. A {\bf 81}, 012304 (2010); A. Bayat, L. Banchi, S. Bose, P. Verrucchi, Phys. Rev. A {\bf 83}, 062328 (2011).
\bibitem{boson} W. S. Bakr, et al., Nature 462, 74 (2009); M. Greiner, et al.,
Nature 415, 39 (2002); W. S. Bakr, et al., Science 329, 547
(2010).
\bibitem{fermion} R. J¨ordens, et al., Nature 455, 204 (2008); U. Schneider, et al.,
Science 322, 1520 (2008).

\bibitem{Lukin} L. Duan, E. Demler and M. D. Lukin, Phys. Rev. Let. {\bf 91}, 090402 (2003).

\bibitem{Meschede} M. Karski {\em et al.}, New J. Phys. {\bf 12}, 065027 (2010).

\bibitem{Weitenberg} C. Weitenberg {\em et al.}, Nature {\bf 471}, 319 (2011).

\bibitem{Gibbons-Nondestructive} M. J. Gibbons, C. D. Hamley, C. Y. Shih and M. S. Chapman, arXiv:1012.1682.

\bibitem{single-site} J. F. Sherson, {\em  et al.}, Nature {\bf 467}, 68 (2010);
C. Weitenberg {\em  et al.}, Nature {\bf 471}, 319 (2011).

\bibitem{supperlattice1} A. M. Rey, et al., Phys. Rev. Lett. 99, 140601 (2007).

\bibitem{Trotzky1} S. Trotzky, P. Cheinet, S. Fölling, M. Feld, U. Schnorrberger, A. M. Rey, A. Polkovnikov, E. A. Demler, M. D. Lukin and I. Bloch, Science {\bf 319}, 295 (2008);

\bibitem{Trotzky2} S. Trotzky, Y. -A. Chen, U. Schnorrberger, P, Cheinet, and I, Bloch, Phys. Rev. Lett. {\bf 105}, 265303 (2010).

\bibitem{supperlattice2} P. Medley, D. M. Weld, H. Miyake, D. E. Pritchard and W. Ketterle, Phys. Rev. Lett. {\bf 106}, 195301 (2011).
\bibitem{clark} S. R. Clark, C. M. Alves and D. Jaksch, New J. Phys. {\bf 7}, 124 (2005).
\bibitem{wooters} W. K. Wootters, Phys. Rev. Lett. {\bf 80}, 2245 (1998).

\bibitem{Leib} E. Lieb, T. Schultz, and D. Mattis, Ann. Phys. (N.Y.) {\bf 16}, 407 (1961).

\bibitem{jordan-wigner} P. Jordan and E. Wigner, Z. Phys. {\bf 47}, 631 (1928).

\bibitem{venuti} L. C. Venuti, S. M. Giampaolo, F. Illuminati and P. Zanardi, Phys. Rev A {\bf 76}, 052328 (2007).

\bibitem{dimer_xy} E.B. Fel'dman, M.G. Rudavets, JETP Letters {\bf 81} 47 (2005); S.I.Doronin, E.B.Fel'dman, A.N.Pyrkov, JETP Letters, {\bf 85}, 519 (2007).
\bibitem{even-xy} E. I. Kuznetsova and E. B. Feld’man, JETP {\bf 102}, 882 (2006).

\bibitem{xxz-phase1} M. Christandl, N. Datta, A. Ekert, and A. J. Landahl, Phys.
Rev. Lett. {\bf 92}, 187902 (2004); M. B. Plenio and F. L. Semiao,
New J. Phys. {\bf 7}, 73 (2005); A. Wojcik, T. Luczak, P. Kurzynski,
A. Grudka, T. Gdala, and M. Bednarska, Phys. Rev. A {\bf 72},
034303 (2005); A. Kay, Phys. Rev. Lett. {\bf 98}, 010501 (2007);
C. Di Franco, M. Paternostro, and M. S. Kim, Phys. Rev. Lett.
{\bf 101}, 230502 (2008).
\bibitem{xxz-phase2} H. Mikeska and A. Kolezhuk, Lect. Notes Phys. 645, 1
(2004).

%%%%%%%%%%%%%%%%%%%%%%%%%%%%%%%%%%%%%%


%\bibitem{Bremner} M.~J. {Bremner} {\em et~al.}, Phys. Rev. Lett. {\bf 89}, 247902 (2002).


%\bibitem{Wilk-Saffman}
%T.~{Wilk} {\em et~al.}, Phys. Rev. Lett. {\bf 104}, 010502 (2010);
%L.~Isenhower {\em et~al.}, Phys. Rev. Lett. {\bf 104}, 010503 (2010).

%\bibitem{Jones}P. H. Jones, M. Goonasekera, and F. Renzoni, Phys. Rev. Lett.
%{\bf 93}, 073904 (2004); R. Gommers et al., Phys. Rev. Lett. {\bf
%94}, 143001 (2005).

%\bibitem{Salger-rachet} T. Salger  {\em et~al.}, Science {\bf 326}, 1241 (2009); S. Denisov, L.
%Morales-Molina, S. Flach, and P. Hnggi, Phys. Rev. A {\bf 75},
%063424 (2007).

%\bibitem{Hanggi-rachet} P. Hanggi, F. Marchesoni, Rev. Mod. Phys. {\bf 81}, 387 (2009).

%\bibitem{Cirac-bus} J.~Cirac and P.~Zoller, Nature {\bf 404}, 579 (2000).

%\bibitem{Bloch-AC-driving} Y.~A.~Chen,  {\em et al.}, arXiv:1104.1833.

%\bibitem{creffield-gate} C. E. Creffield, Phys. Rev. Lett. {\bf 99}, 110501 (2007).

%\bibitem{quantum-rachet} K.~Hai, W.~Hai, and Q.~Chen, Phys. Rev. A {\bf 82}, 053412 (2010);
%O.~R.~Isart, J.~J.~G.~Ripoll, Phys. Rev. A {\bf 76}, 052304
%(2007).



%\bibitem{Mandel-gate} O.~{Mandel} {\em et~al.}, Nature {\bf 425}, 937 (2003).

%\bibitem{Meschede-gate} M.~Karski  {\em et~al.}, J. Korean Phys. Soc. {\bf 59}, 2947 (2011).



%\bibitem{singlet-attach} A. Bayat and S. Bose, Phys. Rev A {\bf 81}, 012304 (2010); A. Bayat, L. Banchi, S. Bose and P. Verrucchi Phys. Rev A {\bf 83}, 062328 (2011).

%\bibitem{rafiee_bayat} M. Rafiee and A. Bayat, arxiv:1105.1809.

%\bibitem{Bloch_swapp} P. Barmettler, A. M. Rey, E. Demler, M. D. Lukin, I. Bloch, and V.
%Gritsev, Phys. Rev A {\bf 78}, 012330 (2008).

%\bibitem{Nakahara} M. Nakahara, T. Ohmi and Y. Kondo,
%arXiv:1009.4426; E. H. Lapasar, K. Kasamatsu, Y. Kondo, M.
%Nakahara, T. Ohmi, J. Phys. Soc. Jpn. {\bf 80}, 114003 (2011).

%\bibitem{Cheinet}P. Cheinet, et al., Phys. Rev. Lett. {\bf 101}, 090404 (2008); S. Trotzky, et al., Phys. Rev. Lett. {\bf 105}, 265303 (2010).

%\bibitem{Nielsen} M. A. Nielsen and I. L. Chuang, Quantum computation
%and quantum information (Cambridge University Press,Cambridge,
%2000).

%\bibitem{Zanardi} P. Zanardi and D. A. Lidar, Phys. Rev. A  {\bf 70}, 012315 (2004);

%\bibitem{Neilsen} M. A. Neilsen, Physics. Letters A {\bf 303}, 249 (2002);

%\bibitem{bollinger}
%J.~Bollinger {\em et~al.}, IEEE Trans. on Instrum. and Measurement
%{\bf 40}, 126 (1991).

%\bibitem{normalized-hopping} F. Grossmann  {\em et~al.}, Phys. Rev. Lett. {\bf 67}, 516 (1991);
%M.~Holthaus, Phys. Rev. Lett. {\bf 69}, 351 (1992); C. E.
%Creffield, Phys. Rev. B {\bf 67}, 165301 (2003).


\end{thebibliography}
\end{document}